\documentclass[12pt,preprint]{aastex}

\newcommand{\etal}{et~al.~}

\newcommand{\msun}{M$_{\sun}$}

\newcommand{\vsini}{$v \sin i$}

\received{2006 September}
\begin{document}

\title{Rotational Velocities For B0-B3 Stars in 7 Young Clusters: Further Study of the Relationship between Rotation Speed and Density in Star-Forming Regions}

\author{S.\ C.\ Wolff, S.\ E.\ Strom, D.\ Dror}
\affil{National Optical Astronomy Observatory, 950 N. Cherry Ave., Tucson, AZ, 85719 (swolff@noao.edu)}

\and
\author{K.\ Venn}
\affil{University of Victoria, Victoria BC V8W 2Y2}

\begin{abstract}

We present the results of a study aimed at assessing the differences in the distribution of  rotation speeds, N ({\vsini}) among young (1-15 Myr) B stars spanning a range of masses 6 $<$ M/{\msun} $<$ 12 and located in different environments: 7 low density ($\rho <$ 1 {\msun}/$pc^{3}$) ensembles that are destined to become unbound stellar associations, and 8 high density ($\rho \gg$ 1 {\msun}/$pc^{3}$) ensembles that will survive as rich, bound stellar clusters for ages well in excess of $10^8$ years. Our results demonstrate (1) that independent of environment, the rotation rates for stars in this mass range do not change by more than 0.1 dex over ages t $\sim$ 1 to t $\sim$ 15 Myr; and (2) that stars formed in high density regions lack the cohort of slow rotators that dominate the low density regions and young field stars. We suggest that the differences in N({\vsini}) between low and high density regions may reflect a combination of initial conditions and environmental effects: (1) the higher turbulent speeds that characterize molecular gas in high density, cluster-forming regions; and (2) the stronger UV radiation fields and high stellar densities that characterize such regions. Higher turbulent speeds may lead to higher time averaged accretion rates during the stellar assembly phase. In the context of stellar angular momentum regulation via ``disk-locking," higher accretion rates lead to both higher initial angular momenta and evolution-driven increases in surface rotation rates as stars contract from the birthline to the Zero Age Main Sequence. Stronger UV radiation fields and higher densities may lead to shorter disk lifetimes in cluster-forming regions. If so, B stars formed in dense clusters are more likely to be ``released"  from their disks early during their PMS lifetimes and evolve into rapid rotators as they conserve angular momentum and spin up in response to contraction. By contrast, the majority of their brethren in low density, association forming regions can retain their disks for much or all of their PMS lifetimes, are ``locked" by their disks to rotate at constant angular speed, and lose angular momentum as they contract toward the ZAMS, and thus arrive on the ZAMS as relatively slowly rotating stars.

\end{abstract}


\section{Introduction}

There is growing evidence that the distribution of rotation speeds of B-type stars depends on the environment in which the stars formed.  More than two decades ago, Wolff, Edwards and Preston (1982; hereafter WEP82) noted that the average rotation speed for B stars in the relatively dense Orion Nebula Cluster (Orion Id in the notation first introduced by Warren and Hesser [1977]) is significantly higher than that found for stars of similar type located in much lower density regions of the Orion star-forming complex (Orion Ia, Ib, Ic).  Moreover, examination of Figure 7 of WEP82 reveals that the B stars in Orion Id lack the cohort of slow rotators ({\vsini} $ < $ 50 km/sec) found in abundance in Orion Ia, Ib and Ic.

At about the same time, Guthrie (1982) studied late B-type stars in 13 clusters and found that the distribution for cluster stars was bimodal with one peak at {\vsini} $ < $ 50 km/sec and a second peak at {\vsini} $\sim $ 225 km/sec, while the distribution of rotational velocities for field stars peaked at {\vsini} $ < $ 50 km/sec and declined monotonically with increasing rotation rates: a result consistent with that of WEP82.

More recently, Strom, Wolff, and Dror (2005) examined the distribution of rotation speeds in the bound galactic clusters h and  $\chi$ Per (age $\sim$ 13 Myr; Slesnick et al. 2002), with the goal of testing whether these very dense regions are, like Orion Id, also characterized by rapid rotation.  From observation of 216 stars in h and  $\chi$ Per spanning types B0 through A0, these authors concluded that: (1) the distribution of projected rotation speeds, N({\vsini}), for stars of types B8-A0, which have evolved little from the Zero Age Main Sequence, exhibits a mean value more than twice that of field star analogs of comparable age; the h and  $\chi$ Per sample also lacks the cohort of slow rotators ({\vsini} $ < $ 50 km/sec) common among field stars, most of which likely originate in low density, unbound OB Associations; (2) the rotation speeds among h and  $\chi$ Per members among earlier spectral types, which have begun to evolve significantly from the ZAMS, are also higher than for their field counterparts; however, the difference in mean rotation speeds is less.  The basic trends observed for h and  $\chi$ Per thus confirm those initially reported in WEP82 for the Orion star-forming complex.  Other recent studies (Huang and Gies 2006a; Dufton et al. 2006) also find fewer slow rotators among the cluster B-type stars relative to nearby B stars in the field.

Two explanations have been offered for the differences in the rotational velocity distributions of the cluster stars as compared with the field stars. The first proposes that the fiducial field star samples are on average older than their cluster counterparts. Such an age difference, combined with the decrease in surface rotation rate expected as a consequence of expansion and/or angular momentum loss via stellar winds could account naturally for the observed patterns of rotation speeds between the two cohorts. 
A second possibility is that the differences in rotational velocities reflect differences in initial angular momenta or early angular momentum evolution related in some way to the environment in which cluster and field stars formed initially.  Current theory provides a plausible framework for explaining the observed trends. Specifically, Konigl (1991) and  Shu et al. (1994)  (see also Paatz and Camenzind 1996) suggested more than a decade ago that initial stellar angular momenta are established during the primary stellar accretion phase via locking of stellar angular velocity to the angular velocity of the circumstellar accretion disk at or near the radius, $r(m)$, where the stellar magnetosphere links to the disk. That radius is set by the balance between the dynamical pressure of accreting material and the magnetic pressure of the magnetic field rooted in the forming star. For a fixed stellar magnetic field strength, the higher the accretion rate through the disk, the smaller the $r(m)$, the higher the Keplerian rotation speed of the disk at $r(m)$, and hence the higher the angular rotational speed of the star. Hence, if accretion rates are higher in the regions that give birth to high density bound clusters, the resulting stellar population would be expected to exhibit higher rotation speeds on average.

Recent work suggests that accretion rates during the stellar assembly phase may indeed be higher in regions that give birth to stellar clusters.  McKee \& Tan (2003) and Elmegreen \& Shadmehri (2003) both note that dense stellar clusters form in molecular complexes characterized by very high gas surface density, as well as by close packing of protostars and star formation efficiencies high enough to ensure the formation of a bound cluster. These authors argue that the turbulent velocity of the gas in these regions is likely to be high as well, thus requiring that star-forming cores have high initial densities in order that self-gravity exceed internal thermal plus turbulent pressure. High initial densities in turn lead to short protostellar collapse times ($t_{collapse} \sim \rho^{-1/2}$) and as a consequence, high time-averaged accretion rates during the stellar assembly phase.

In this contribution, we extend our study of the relationship between observed rotation speeds and density to six young clusters and OB associations (ages 1-5 Myr).  Our goals are first to establish whether the distribution of stellar rotation speeds in dense, bound clusters differs significantly from that characterizing unbound associations and the field, and second to understand whether these differences reflect the effects of stellar and/or angular momentum evolution or initial conditions. We focus here on stars in the spectral type range B0.5-B3 primarily because (1) such stars are intrinsically luminous and can be observed to large distances, thus opening the possibility of examining rotational velocity distributions in extragalactic star-forming regions characterized by even higher densities than h and  $\chi$ Per; (2) we expect that the effects of winds and internal mixing on the evolution of surface rotation speeds will be modest compared to their more luminous O-star counterparts; and (3) the role of macroturbulent line broadening in these stars is also negligible in comparison with O stars.  We will combine our new results for these six clusters and associations with results in the literature for stars in the same mass range that are members of the expanding associations Sco-Cen, I Lac, and Orion Ia, b, and c.  

The combined sample of newly-measured rotation speeds and extant published results allows us to examine evolutionary and environmental trends for stars with a range of ages from 1-15 Myr drawn from both dense, bound and low density, unbound groups of stars.  Our approach will be (1) to examine the observed surface density of stars in each of the sample clusters and associations to assess whether their constituent stars formed (a) in relatively long-lived clusters that are likeliest to form in regions characterized by high star-forming efficiency, high stellar density, and high turbulent speeds, or (b) in expanding associations, formed in regions characterized by lower star-forming efficiency, lower density and lower turbulent speeds; such stars are likely over time to join the population of field B stars;  (2) to assess the magnitude of evolutionary changes among stars of different ages but born in regions of similar density; and (3) to compare the distribution of apparent rotation speeds, N({\vsini}) for stars born in bound clusters and unbound associations.

\section{The Samples}

The previous studies cited above, as well as many others, have established that the distribution of rotational velocities for early B stars in any given cluster or association is broad, with individual stars having projected rotational velocities as low as a few km/sec and as high as $\sim$ 400 km/sec.  Therefore in order to establish whether or not there are statistically significant differences in the rotational velocity distributions of stars of different ages or in different environments, we need fairly large samples.  The studies of galactic OB associations by Massey \& Thompson (1991), Massey, Johnson, \& DeGioia-Eastwood (1995), and Hillenbrand \etal (1993) provide good source lists along with photometry and spectral classification of early type stars.  We have obtained new values of {\vsini} for early B stars in several of these associations and have supplemented our new observations with previously published values for members of the nearby expanding associations in Orion, Scorpio-Centaurus (Sco-Cen), and Lacerta.  Unbound associations are presumably the types of regions that give rise to field stars, but in contrast to individual field stars, an accurate age can be estimated from the H-R diagram for each association.

\section{New Observations and Data Reduction}

The B stars for which we obtained new spectroscopic observations were members of NGC 6823, NGC 7380, IC 1805, NGC 2244, NGC 6611, and Cyg OB2 and were selected on the basis of observed colors, spectral types, and membership studies by Massey \& Thompson (1991), Hillenbrand \etal ~(1993), and Massey, Johnson, \& DeGioia-Eastwood (1995).  Our primary selection criterion is color. To be included in our sample, a star's unreddened $Q$ color must fall within the range $-0.86 < Q < -0.46$, where $Q = (U-B) - 0.72(B-V)$; this color range corresponds approximately to the spectral type range B0.5 to B3. For NGC 6611 only, observations were obtained in 2005 May, with the Hydra multiobject fiber spectrograph on WIYN.  In conjunction with an echelle grating and an order separating filter, these observations yielded spectra with R $\sim$ 20,000 spanning the wavelength range 4450 to 4590\AA.  Within this wavelength range, the only line useful for measuring rotation for early B stars is He I 4471\AA, for which we measured the full-width half-maximum (FWHM).

For many of the same stars in NGC 6611 and for the stars in the other associations, we obtained WIYN-Hydra observations that cover the wavelength range 4070-4580\AA~at a resolution of R $\sim$ 5000 (0.75\AA).  These observations enable measurement of projected rotational velocities, {\vsini} $>$ 50 km/sec, and an upper limit of 50 km/sec for stars rotating more slowly.  The spectra were extracted and calibrated in wavelength by Kim Venn.

For the lower resolution (R $\sim$ 5000) data, values of projected rotational velocities were obtained from measurement of FWHM for He I 4387\AA~and 4471\AA, with the occasional inclusion of Mg II 4481\AA~when its strength was comparable to that of the helium lines.  It is important to note that the FWHM of He I 4471\AA~is sensitive not only to rotational velocity but also to gravity for stars in the spectral type range B0.5 to B3.  For example, models by Daflon \& Cunha (private communication) indicate that at $T_{eff}$ = 25000 K, a calibration of {\vsini} in terms of FWHM for stars characterized by log g = 3.5 is offset by 50 km/sec from the calibration for stars having log g = 4.5.  Hence, in deriving {\vsini} values from measured FWHM, it is essential that sample and calibrator stars have comparable surface gravities. With the exception of NGC 6823, in which the stars have a range of ages from 2-7 Myr, the stars in all of our clusters are estimated to be less than 5 Myr old (Massey \etal ~1995).  These young stars have typical surface gravities in the range log g $\sim$ 4.0-4.2 (Dufton \etal ~2006; Huang \& Gies ~2006b).  We have used 18 stars in two of these young clusters (IC 1805 and NGC 2244) as calibration stars. Values of {\vsini} for this cohort were derived by Huang \& Gies (2006a), who used the stellar atmosphere code TLUSTY and the radiative transfer code SYNSPEC (Hubeny \& Lanz 1995) to construct a grid of helium line profiles for surface gravities appropriate for their nominal age (t $\sim$ 1-3 Myr) and for a range of rotational velocities.   We have used the values of {\vsini} derived from their analysis to establish a relationship between FWHM and {\vsini} for our data.

Figure 1 compares the values we have derived with those of the calibrators and with measurements of other large samples:  NGC 6611 stars by Dufton \etal (2006), NGC 1805 and NGC 2244 by Huang \& Gies (2006a), and NGC 2244 by Verschueren (1991) and reported on WEBDA (available at http://www.univie.ac.at/webda/webda.html).  The standard deviation of a single measurement from a 45 degree straight line is 27 km/sec.  This uncertainty compares favorably with the value of 35 km/sec derived from a comparison of values quoted by Huang \& Geis (2006a) and Dufton \etal (2006) respectively for main sequence B1-B3 stars in NGC 4755 and NGC 3293; both of these studies used model atmospheres and profile fitting to derive the values of {\vsini}.  We conclude that our derived values of {\vsini} are both calibrated to a scale established from comparison with models and have comparable uncertainties for a single measurement.

Our results are summarized in Table 1, which lists for each star in our sample: the Webda number, or in the case of Cyg OB2, the number from Massey \& Thompson (1991), position, spectral type, if available, and photometry from Massey \& Thompson (1991), Hillenbrand \etal (1993), or Massey \etal (1995), and our value of {\vsini}.  Given the generally good agreement between our values and those of Huang and Gies and Dufton \etal we have also included in Table 1 their measurements of {\vsini} for those stars in IC 1805, NGC 2244, and NGC 6611 that fall within the specified color range and which we did not observe.  

\section{Extant Observations for Additional Clusters and Associations}

In order to increase the sample of rotational velocity measurements for stars located in low density environments, we have supplemented our newly obtained {\vsini} values with data already available in the literature for several nearby associations: I Lac, Sco OB2, and Orion I.   For the three subgroups in Sco OB2 (Upper Scorpius, Upper Centaurus-Lupus, and Lower Centaurus Crux), we have taken the membership data from the Hipparcos census by deZeeuw \etal (1999).  UBV photometry is not available for all of these stars.  Therefore, we have used the relationships between $Q$ and $T_{eff}$ given by Massey \etal and Lyubimkov \etal (2002) to estimate that $-0.46 > Q > -0.86$ corresponds approximately to 4.17 $<$ log $T_{eff} <$ 4.45, and have used the values of log $T_{eff}$ from deGeus, deZeeuw, \& Lub (1989) to determine which stars to include in our sample.  We have taken the rotational velocities for these stars primarily from Brown \& Verschueren (1997), supplemented by some additional data available from the Simbad databases for stars in Upper Sco.

For I Lac, we have again taken the membership list from deZeeuw \etal (1999).  Most of the listed rotational velocities come from the compilation of Abt, Levato, \& Grosso (2002); the remainder are from the Simbad database.  We included all the B1-B3 stars as well as those stars for which later spectral types are quoted, but which have UBV colors from the Simbad database that place them within the specified color range $-0.46 > Q > -0.86$ that defines the limit of our sample.

The Orion a, b, and c associations are also well studied, and the data on members have been taken from Brown, deGeus, \& de Zeeuw (1994).  Because this paper provides effective temperatures for stars rather than UBV colors, we have again selected those with 4.17 $<$ log $T_{eff} <$ 4.45.  Orion d has too few B stars with {\vsini} values to yield a sample of interesting size and is not included here. 

As an example of old, dense clusters, we have included rotational data for h and  $\chi$ Persei from Strom, Wolff, \& Dror (2005), again limiting our sample to stars with $-0.46 > Q > -0.86$ according to the photometry from Slesnick, Hillenbrand, \& Massey (2002).

We list in Table 2 the star designations and the values of {\vsini} culled from the literature for stars located in Sco OB2, I Lac, and Orion.  

While the data in the literature are somewhat heterogeneous and may not cover exactly the same mass range as the data for the associations for which Massey \etal obtained photometry, the differences are likely to be small.  Fortunately observations have shown that both the average values and the distributions of {\vsini} vary slowly with mass (Abt, Levato, \& Grosso 2002; Huang \& Gies 2006a) for B stars, and any slight differences in the mass boundaries established for each region will not affect our analysis.  There is also sufficient intercomparison among the various rotation papers to show that their velocities are on the same scale to within $\sim$ 10\%. 

\section{Identifying Candidate Bound Clusters and Unbound Associations}

In order to characterize the environments in which the stars included in our sample formed, we list in Table 3: 1) the name of the parent cluster/association; 2) its distance; 3) its age; 4) the total number of early B stars in our specified mass range from the membership lists obtained as described above (note that {\vsini} has not been measured for all of the stars in the defined mass range; that is the explanation for the fact that the numbers of stars in Table 3 do not match the numbers listed in Tables 1 and 2); 5) the median distance in degrees of the B stars in our sample from the center of the cluster; the location of the center is estimated by calculating the average position of the ensemble of B stars; 6) the median radius in pc; 7) an estimate of the total mass of the cluster;  and 8) an estimate of the density of the cluster.

The distances for the subgroups in Sco OB2 and I Lac are from deZeeuw \etal (1999); the ages for the Sco OB2 subgroups are from de Geus, deZeeuw, \& Lub (1989) and for I Lac from Blaauw (1964); for Orion a, b, and c the ages and distances are from Brown \etal (1994);  for h and  $\chi$ Per from Slesnick, Hillenbrand, \& Massey (2002); for Cyg OB2 from Massey \& Thompson (1991); for NGC 6611 from Hillenbrand \etal (1993); and for the remaining associations from Massey \etal (1995). 

One of the advantages of the mass range we have chosen here (approxmately 6-12 {\msun} according to the model atmosphere analyses of Lyubimkov \etal 2002, or approximately spectral types B0.5-B3) is that the lists of members in this mass range should be essentially complete because these stars are among the brightest in the clusters and associations surveyed.  Moreover, the level of contamination from foreground or background objects is expected to be insignificant as well. We further expect that all of the stars formed initially in this mass range will still be within the main sequence band in the HR diagram.  The estimated main sequence lifetime for a 12 {\msun} star is 16 Myr (Schaller, Schaerer, Meynet, \& Maeder 1992) and for a 9 {\msun} star is 26 Myr; all of our associations and clusters are younger than 16 Myr, and most are less than half this age. 

To calculate the total mass of the cluster, we have assumed that the stars in the mass range 6-12 {\msun} constitute 5.5\% of the total mass of the complete population of stars spanning the range 0.1-100 {\msun}.  This is approximately the fraction derived whether we use the Salpeter (1955) mass function or the Miller \& Scalo (1979) function. One check on the robustness of our estimate is provided by the study of h and  $\chi$ Per (Slesnick, Hillenbrand, \& Massey, 2002) who estimate the total mass of the stars with $M >$ 1 $M_{\odot}$ in h and  $\chi$ Per to be 3700 and 2800 {\msun}, respectively.  According to both the Salpeter and Miller-Scalo mass functions, about 60\%  of the mass is contributed by stars with $ M<$  1 {\msun}.  If the Slesnick \etal estimate is extended to 0.1 {\msun}, then their estimated masses become 9200 and 7000 respectively or about 60\% of the values we derive here based on the number of stars in the range 6-12 {\msun}. Another check is provided by the work of Weidner \& Kroupa (2006) who have used the same photometric data as we have for NGC 2244 and NGC 6611 to estimate total masses of 6240 {\msun} and 20,000 $\pm$ 10,000 {\msun} for these two associations, respectively.  Our values are 7800 and 25000.  If we apply our method to Orion d, we estimate the mass in stars to be 2000 {\msun}, which compares well with the value of 1800 {\msun} derived by Hillenbrand \& Hartmann (1998) from star counts.  These estimates suggest that extrapolations from a small number of early type stars to total mass can lead to answers differing by no more than 40\%, presumably a result of the combined effects of Poisson statistics, small differences in  the adopted relationship between colors and mass, and estimates of incompleteness as a function of magnitude.

In order to identify candidate bound clusters and unbound associations, we make use of the definition for bound clusters adopted by Lada and Lada (2003): stellar ensembles comprising more than ~35 members and having a density exceeding 1 {\msun}/$pc^3$. This definition rests on the classic work of Spitzer (1958) who showed that the lifetime of a stellar ensemble is inversely proportional to its density, under the assumption that tidal interactions with interstellar clouds is the primary mechanism for disrupting the cluster. Specifically, $t_{dissipation} \sim  2 \times 10^8 ~~\rho_{cluster}$ years, where $\rho_{cluster}$ is expressed in units of solar mass/$pc^3$. Examination of the estimated densities tabulated in column 8 of Table 3 shows a natural division between candidate bound clusters (with typical densities ranging from 4-400 {\msun}/$pc^3$, and hence survival times $t \gg 10^8$ years), and unbound associations, with typical densities ranging from 0.05-0.7 {\msun}/$pc^3$ and corresponding survival times from a few tens of millions of years to $\sim 10^8$ years.

\section{Analysis}

\subsection{Evolutionary Effects}

The first step in the analysis is to determine whether there are any systematic changes in rotation as stars age.  If there are, then stars must be closely matched in age before we can evaluate environmental effects.  In order to test for evolutionary effects, we divided the stars into four groups:  young, low density; old, low density; young, high density; and old high density.  The cumulative distributions of rotational velocities for samples matched in density but different in age are shown in Figures 2 and 3.  The age difference between the young and old stars in the low density regions is from 2-6 Myr for the young stars to 11-16 Myr for the old stars.  Typical ages for the young stars in the high density regions are 1-5 Myr, while the older stars are from h and $\chi$ Per and have an age of nearly 13 Myr.
A K-S test indicates that the probability that the distributions for the two different age groups of low density associations are drawn from the same parent sample is 0.15; for the two age groups of high density regions, the probability is 0.02.  In other words, the differences in the distributions of the stars in low
density regions are too small to be significant, and the differences in the two
age groups of stars in high density regions are only marginally significant.  Furthermore, even if the small differences in distributions are interpreted as real, the direction of rotational evolution is different in the two cases:  the stars in the low density regions appear to speed up slightly, while the stars in the high density regions appear to slow down with increasing age.  Therefore, we conclude that there is little evidence for evolutionary changes in rotation in the mass range observed here. If we accept the difference shown in Figure 3 for the two age groups of stars in high-density regions as real, then we can estimate an upper bound of $\sim$ 0.1 dex for the change in rotational velocity over an age range of 12 Myr.

Our conclusion that evolutionary effects are small in this mass range is in agreement with previous observational studies by Huang \& Gies (2006a, b) and with models by Meynet \& Maeder (2000) and Heger \& Langer (2000).  In order to look for evolutionary effects, Huang \& Gies (2006a) divided their sample of early-type stars drawn from 19 clusters into two groups.  There was no evidence of rotational evolution for stars in the first group, those with masses $M <$ 9 {\msun}.  For the higher mass stars in the second group, which extended to spectral type O9 and therefore to somewhat higher mass than our study, a comparison of cumulative distributions for different age groups also showed no evidence for rotational evolution.  In a subsequent study, Huang \&Gies (2006b) assigned ages to individual stars in their cluster sample, and for their group with 8.5 {\msun} $< M<$ 16 {\msun} found only a modest change in mean {\vsini} from 134 km/sec for stars near the ZAMS to 106 km/sec for stars nearing the terminal age main sequence (TAMS), a result in approximate agreement with that reported above for the high density regions.  The lowest mass treated by the two sets of model calculations cited above is 12 {\msun}, comparable to the highest  mass stars in our B0.5-B3 cohort. For a star of 12 {\msun}, both groups predict that a ZAMS star rotating at about 200 km/sec will be rotating about 25\% more slowly or at about 150 km/sec when it reaches the TAMS.  Furthermore, the percentage change in rotational velocity is nearly independent of the initial value.  For a large sample, the
average rotational velocity $<v> = (4/\pi)<vsini>$ and so the predicted 
percentage
change in \emph{v} and {\vsini} are the same.  Therefore, in Figure 3, which is a logarithmic plot, we would expect to see a constant shift of 0.1 in the log, which is approximately consistent with the data for the stars in dense regions.

As we shall see, an evolutionary effect of this magnitude, even if real, would not account for the differences in the {\vsini} distributions of stars in high and low density environments.  For the subsequent analysis, we will therefore group all of the stars in low density environments, independent of age, to compare with the stars formed in high density environments, again grouped without consideration of age.

\subsection{Environmental Differences}

In Figure 4 we plot the cumulative velocity distributions for stars formed in low and high density environments.  A K-S test indicates that the probability that these two datasets are drawn from the same parent sample is 0.001.  The difference between the distributions for the stars formed in high density environments and field stars is even larger as shown in Figure 4.  The chances that these two distributions are drawn from the same parent distribution are less than 0.00005.  In other words, the differences in the distributions for the stars in the
three different types of regions are highly significant and are unlikely to have
occurred by chance.  In order to establish the field star sample, we have used the data for single stars from Abt \etal (2002) as tabulated by Strom \etal (2005);  we have used the calibration between the Stromgren index $c_1$ and $T_{eff}$ as given by Lyubimkov \etal (2002) to limit the field star data to the same temperature range as the data on stars in dense associations. Figure 5 shows the histograms for stars in the three different types of environments.  Note that the primary difference is the large number of slowly rotating stars ({\vsini} $<$ 50 km/sec) in the low density regions 

A number of other studies are consistent with this result in that they find that clusters lack the large number of stars with  {\vsini} $<$ 50 km/sec that are common among field stars.  Because the field stars are not currently members of clusters, they must have formed in low-density, unbound regions.  Recent examples include the study by Keller (2004), who found that cluster stars in the LMC rotate more rapidly than LMC field stars; the survey by Huang \& Gies (2006a), who reported that only about 17\% of the stars in their 19 clusters had {\vsini} $<$ 50 km/sec as compared with the 30\% of field B stars with  {\vsini} $<$ 50 km/sec (Abt \etal 2002); and the study of massive stars by Dufton \etal (2006), who found that fewer than 10\% of the stars in NGC 3293 and NGC 4755 had apparent rotational velocities less than 50 km/sec, as compared with 40-50\% of a comparable field star sample. 

The systematic differences in the cumulative distributions shown in Figure 4 persist to high values of {\vsini} but become progressively smaller until the distributions converge near a maximum value of $\sim$350 km/sec.  This is illustrated very nicely by the independent data of Dufton \etal (2006) for NGC 3293 and NGC 4755.  Because the critical velocity for stars in the mass range surveyed here exceeds 500 km/sec (McSwain \& Gies 2005), gravity darkening is likely to play only a modest role in reducing the apparent line-broadening (Townsend, Owocki, \& Howarth 2004) and in any case was taken into account in the survey by Huang \& Gies (2006a), which we relied on for calibration.  Therefore it appears likely that nearly all early B stars rotate at rates significantly less than critical and that the upper limit on rotation is nearly independent of environment.  This is consistent with the study of Be stars by McSwain \& Gies (2005), who found no correlation between the fraction of Be stars and cluster density.

While we have examined {\vsini} distributions only for early B stars in the current paper, other studies show that the more rapid rotation of cluster stars is characteristic of cooler B stars as well (Dufton \etal 2006; Strom \etal 2005)

\subsection{Role of Initial Conditions}

Following the arguments put forth by Strom, Wolff, \& Dror (2005), we suggest that the observed differences in the distribution of projected rotation speeds among early B-stars in high density, bound clusters and low density unbound associations could reflect differences in the initial and/or environmental conditions that characterize cluster- and association- forming regions.  In the discussion that follows, we focus on the formation of single stars. While one might also expect the formation of binaries to depend on environment, there are not yet either enough data or sufficiently well developed models to examine the issue of binary formation, its dependence on environment, or its effect on stellar rotation.

There appear to be two significant differences between regions that form bound clusters and unbound associations. First, the surface density of cluster-forming molecular clumps is $\Sigma \sim$  1~gm/cm$^2$. In contradistinction, regions such as the Taurus-Auriga or Orion A molecular clouds, which harbor widely distributed populations of forming stars that are likely to disperse as natal molecular material dissipates, exhibit much lower surface densities ($\Sigma \sim$  0.03~gm/cm$^2$; McKee \& Tan 2003). The assertion that regions of high surface density form bound clusters finds strong support from the work of Plume \etal (1997) who note that the observed bolometric luminosity ($L_{bol}$) to total mass ($M_{clump}$) ratio in high surface density clumps implies star-forming efficiencies  SFE $\sim$ 50\%, comparable to the SFE required for emerging stellar groups to remain bound following the dissipation of natal molecular material (see, for example, Lada, Margulis, \& Dearborn 1984); lower surface density clumps are by contrast characterized by much lower values of $L_{bol}/M_{clump}$, suggesting much lower SFEs ($\sim$ 5\%), too small to produce gravitationally bound ensembles. Second, the higher surface density, putatively cluster-forming ‘clumps’ appear as well to be characterized by higher turbulent speeds (e.g. Plume \etal 1997).  Turbulent speeds in such clumps can reach values of 5-10 km/sec as compared with the values of 1-2 km/sec characteristic of the low surface density molecular clouds that appear to form loose associations (e.g. Taurus-Auriga). 

In order to relate environmentally-driven differences in initial conditions to stellar rotation, we need a model for the star formation process.  There is growing evidence (see, for example, Cesaroni \etal 2006; Beuther \& Shepherd 2005; and references therein) that stars in the mass range studied here (up to 12 {\msun}) are formed in the same way as lower mass stars:  from infalling material located in a rotating protostellar core channeled starward through a circumstellar accretion disk, with the flow of material from the inner regions of the accretion disk toward the stellar surface mediated by a magnetic field rooted in the star.  We will refer to this process as magnetically mediated accretion (MMA).

Transporting a large fraction of the mass of a forming star through an accretion disk transports angular momentum as well. As a consequence the forming star should quickly spin up to near breakup rotational velocity.  The prediction that forming stars should be rotating at breakup  (Durisen \etal 1989) is not, however, borne out by observations of either T Tauri stars or intermediate mass stars, most of which are rotating an order of magnitude or so below the breakup velocity.  The classic paper that attempted to reconcile the predicted near-breakup speeds of newly-formed stars with observation is by Konigl (1991), who proposed that a significant spin-down torque could be applied to the star if there were a magnetic connection between the star and disk.  Various modifications of this idea have been proposed (see, for example, Shu \etal 1994; Long, Romanova, \& Lovelace 2005).  Matt \& Pudritz (2005a) have argued that the steady-state disk-locked model described in the original paper by Konigl cannot be correct because differential rotation between the disk and star will cause the field lines to open.  This reduces the spin-down on the star and results in faster rotation than predicted by the steady-state model.  As an alternative Matt \& Pudritz (2005b) have argued that a stellar wind can carry away sufficient angular momentum to ensure that the accreting star rotates slowly.

Whatever the details of the process, the papers by Konigl (1991), Shu \etal (1994), Long \etal (2005), and Matt \& Pudritz (2005b) all yield the same dependences for the rotational velocity;

	$v \sim  M^{5/7}~~\dot{M}_{acc}\,^{3/7}~~B^{-6/7}~~R^{-11/7}$

\noindent where M is the mass of the star, $\dot{M}_{acc}$ is the accretion rate, B is the magnetic field strength, and R is the radius of the star.  Wolff \etal (2004) have shown that this relation with reasonable assumptions for $\dot{M}_{acc}$ and B can account for the trend in the upper bound of angular momentum per unit mass, J/M, for stars in the mass range 0.1-10 {\msun}.  Using somewhat different parameterizations, Long \etal (2005) and Matt \& Pudritz (2005b) indicate that the typical rotational velocity should be about 0.2 times breakup.  Observations show that the median rotational velocities of stars in the mass range 0.1-50 {\msun} are indeed in the range 0.1- 0.2 times the breakup speed (Wolff \etal 2006).

If we accept this model for the formation of stars in the mass range 6-12 {\msun} surveyed here, then the variables that might influence rotation rates in different environments are (a) the magnetic field strength; (b)the accretion rate; and (c) the length of time over which stars contracting toward the main sequence are  kept from spinning up as consequence of  angular momentum transfer from the star to the accretion disk and/or to an accretion driven disk or stellar wind. Since we have no information about magnetic field strengths in early B stars, we will focus on the other two possibilities: accretion rates and disk lifetimes.

There is theoretical justification for the suggestion that accretion rates may be higher in dense environments.  As noted above, cluster-forming molecular clumps appear to have higher turbulent speeds (Plume \etal 1997). In turn, higher turbulent speeds require that individual star-forming cores within the clumps have higher initial densities in order that their self-gravity overcome the higher turbulent pressures characterizing the cores. As noted in the introduction, higher initial core densities lead to shorter collapse times and higher time-averaged accretion rates during stellar assembly (McKee \& Tan 2003). Given that the intial rotation speed on the birthline varies directly with $\dot{M}_{acc}\,^{3/7}$, higher time-averaged accretion rates are expected to lead to higher initial stellar rotation speeds in cluster-forming clumps.

A second factor, which could play a major role in the determining the rotation rates of low mass stars, is the length of time over which the disk survives.  In low mass stars, there is evidence that disk lifetimes can range from no more than a few hundred thousand years to several million years (Haisch, Lada, \& Lada 2001).  The cohort of rapidly rotating ({\vsini} $\sim$ 200 km/sec) solar-like stars among main sequence stars are thought to be the descendants of the pre-main sequence stars that were released early from interaction with their natal accretion disks, and then conserved angular momentum, and spun up as they completed their contraction to the ZAMS.  By contrast, the cohort of slowly rotating main sequence stars ({\vsini} $\sim$ 10-30 km/sec) are thought to be the descendants of stars that continued to lose angular momentum through interaction with an accretion disk until their contraction was nearly completed (see, for example, Rebull, Wolff, \& Strom 2004; Herbst \& Mundt 2005). Environment may play a crucial role in determining disk lifetimes and hence outcome distributions of rotation speeds. More specifically, the environment of a rich cluster is particularly hostile to the long term survival of disks. First, rich clusters are the regions most likely to have significant numbers of closely packed O stars (Weidner \& Kroupa, 2006) that can contribute to rapid disk erosion via photoevaporation (Johnstone, Hollenbach, and Bally 1998; Shen \& Lou 2006). Second, gravitational interactions in such dense clusters can be effective in disrupting disks through tidal stripping (Pfalzner, Olczak, \& Eckart 2006). The sense of both effects is to reduce disk lifetimes in cluster-forming environments, and as a result to produce a larger cohort of more rapidly rotating stars. 

It is important to note, however, that the pre-main sequence evolutionary time scales for the intermediate mass stars studied here are very short.  For example, a 7 {\msun} star takes only about 500,000 years (Siess, Dufour, \& Forestini 2000) to evolve from the stellar birthline to the ZAMS. If these stars are formed with accretion rates similar to those that characterize the main assembly phase for low mass stars ($10^{-5}$ {\msun}/yr; Palla \& Stahler 1992), then the accretion disks for the stars in our sample must survive until the star completes its contraction to the main sequence simply in order to build the star to the observed mass.  The detailed modeling by Palla \& Stahler, for example, shows that at this accretion rate, stars with $M \sim$ 8 {\msun}  are already on the main sequence when the accretion ends.  After the star reaches the main sequence, contraction is completed, and no further spin up is possible after disruption of the disk.  

If, on the other hand, the accretion rate is higher in cluster-forming regions and a 7 {\msun} star is formed at an accretion rate of $10^{-4}$ {\msun}/yr, then it will contract by about a factor of four from its initial radius following its depostion on the stellar birthline at the end of the main accretion phase, to its final radius on the ZAMS (Siess \etal 2000; see also Norberg \& Maeder 2000; Behrend \& Maeder 2001; McKee \& Tan 1993).  While we do not know the rotation rate for 6-12 {\msun} stars near the birthline, we do have data for stars of somewhat lower mass ($M \sim$ 2 {\msun}) shortly after they reach their birthline and are deposited high on convective tracks (Wolff \etal 2004); their apparent rotational velocities range from 10-50 km/sec.  Lower mass ($M < $ 2 {\msun}) T Tau stars on convective tracks, typically rotate at about 0.1-0.3 breakup (Stassun \etal 1999).  If the more massive stars formed in clusters also emerge from the main accretion phase rotating at 0.1-0.3 breakup but well above the ZAMS (as expected for accretion rates $\sim 10^{-4}$ {\msun}/yr) then a factor of four spin up would be sufficient to leave very few stars rotating at less than 50 km/sec, as is observed for stars in rich clusters and associations.

In summary, in the context of MMA, the cohort of slowly rotating B stars in the mass range 6-12 {\msun} must have retained an interaction with their accretion disk essentially until they completed their contraction to very close to the main sequence, either because they were formed from cores characterized by low time-averaged accretion rates, or because the continued presence of a disk kept them from spinning up as they contracted to the main sequence.  The absence of slow rotators in dense, cluster-forming regions would then, via this reasoning, result from a combination of the following factors:  1) formation from cores characterized by higher accretion rates, producing higher initial rotation rates when the main accretion phase ceases; and 2) contraction from larger radii at their initial positions on the stellar birthline, resulting in greater spin up as they contract to the main sequence; or 3) rapid photoevaporation or tidal dissipation of a disk, thereby terminating ``disk locking" during contraction toward the ZAMS.

\section{Summary}

In this paper we have shown that the distributions of {\vsini}, N({\vsini}), for B stars depend on environment, with stars in rich clusters rotating on average more rapidly than stars formed in expanding associations and field stars.  The primary difference between the N({\vsini}) distribution is found among stars with {\vsini} $<$ 50 km/sec:  rich clusters lack the large cohort of very slowly rotating stars found in the field.  We have also shown that the observed values of {\vsini} for stars in the mass range 6-12 {\msun} change very little over the age range spanned by our survey, which is about 12 Myr.  Earlier studies had speculated that the differences between the cluster stars and the field stars might be attributed to systematic differences in age and loss of angular momentum as the stars aged, but we find that this is not the case.

We propose instead that stellar ensembles that survive as bound clusters likely form in high surface density clumps, characterized by high star-forming efficiencies, high turbulent speeds, short core collapse times, high time-averaged accretion rates during stellar assembly, with a consequent bias toward high rotation speeds. Early disruption of disks via photoevaporation or tidal stripping may also be a factor in producing rapidly rotating stars in regions that give birth to stellar clusters. By contrast, the tendency to form a large cohort of slow rotators among B stars born in regions destined to evolve into unbound associations may well reflect the lower rotation speeds that would result from (a) lower surface densities and lower turbulent speeds, and lower time-averaged accretion rates characterizing such regions, along with (b) the fact that the disks around these stars are less likely to experience early disruption through interactions with neighbors or radiation from nearby O stars.

Whether or not these proposed explanations for the observations are correct, our basic result: that B stars in bound clusters lack the cohort of slow ({\vsini} $<$ 50 km/sec) rotators characteristic of their brethren in low density, unbound associations,  appears to require a causal link between initial conditions in star-forming regions and initial stellar angular momenta. Understanding this link would appear essential to developing a comprehensive understanding of the factors which control not only stellar angular momenta, but perhaps other critical properties (e.g. the initial mass function) in different star-forming environments.

\acknowledgements 
We thank Diane Harmer of NOAO, who provided generous assistance both in preparing for our WIYN-Hydra observing run and at the telescope, and P. Dufton, D. Gies and W. Huang for providing data in advance of publication.

\clearpage
\begin{figure}
\plotone{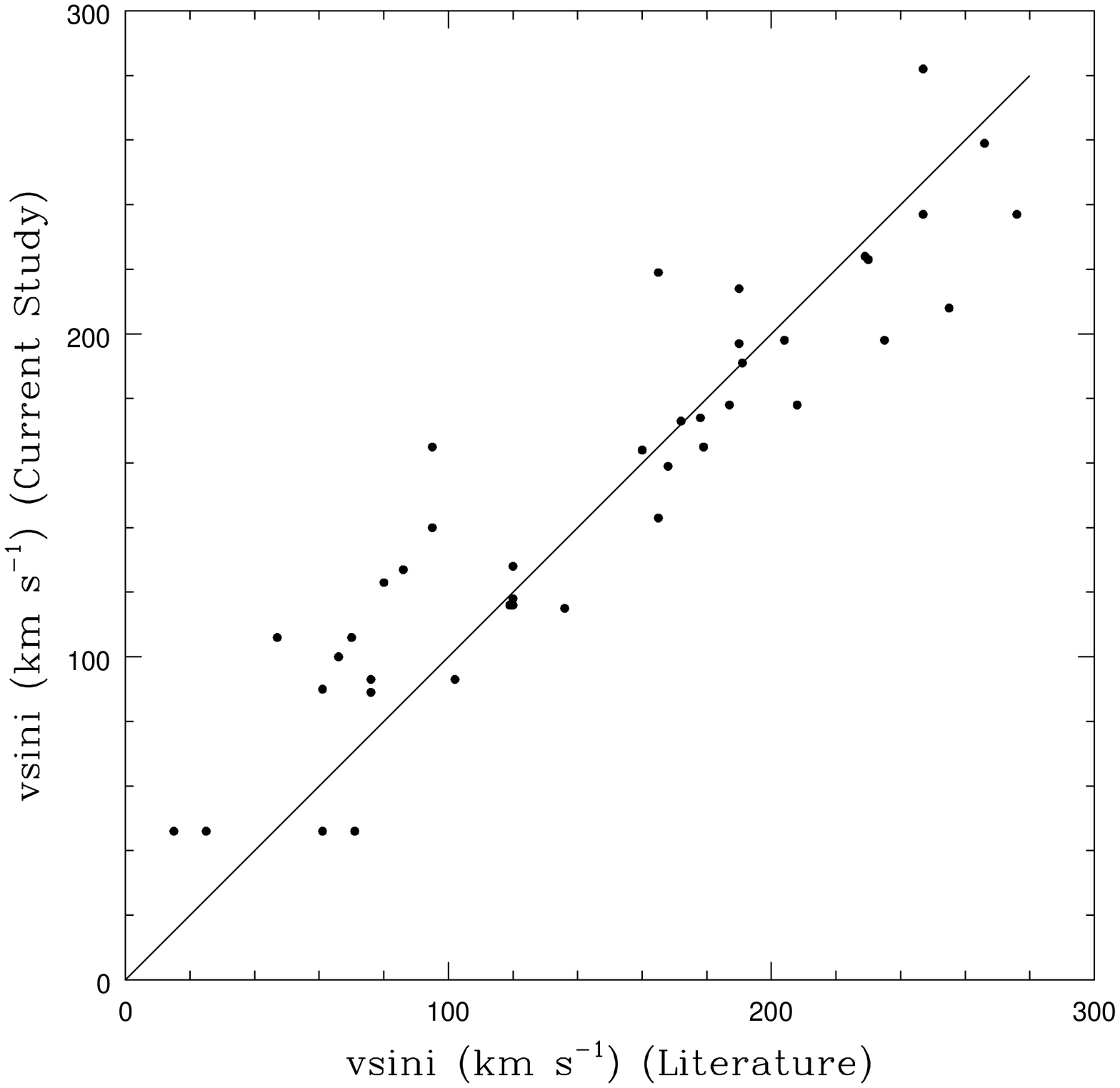}
\figcaption{Plot comparing the values of {\vsini} derived in the current study with those available in the literature.  A 45$^{\circ}$ straight line is also shown.}
\end{figure}
\clearpage
\begin{figure}
\plotone{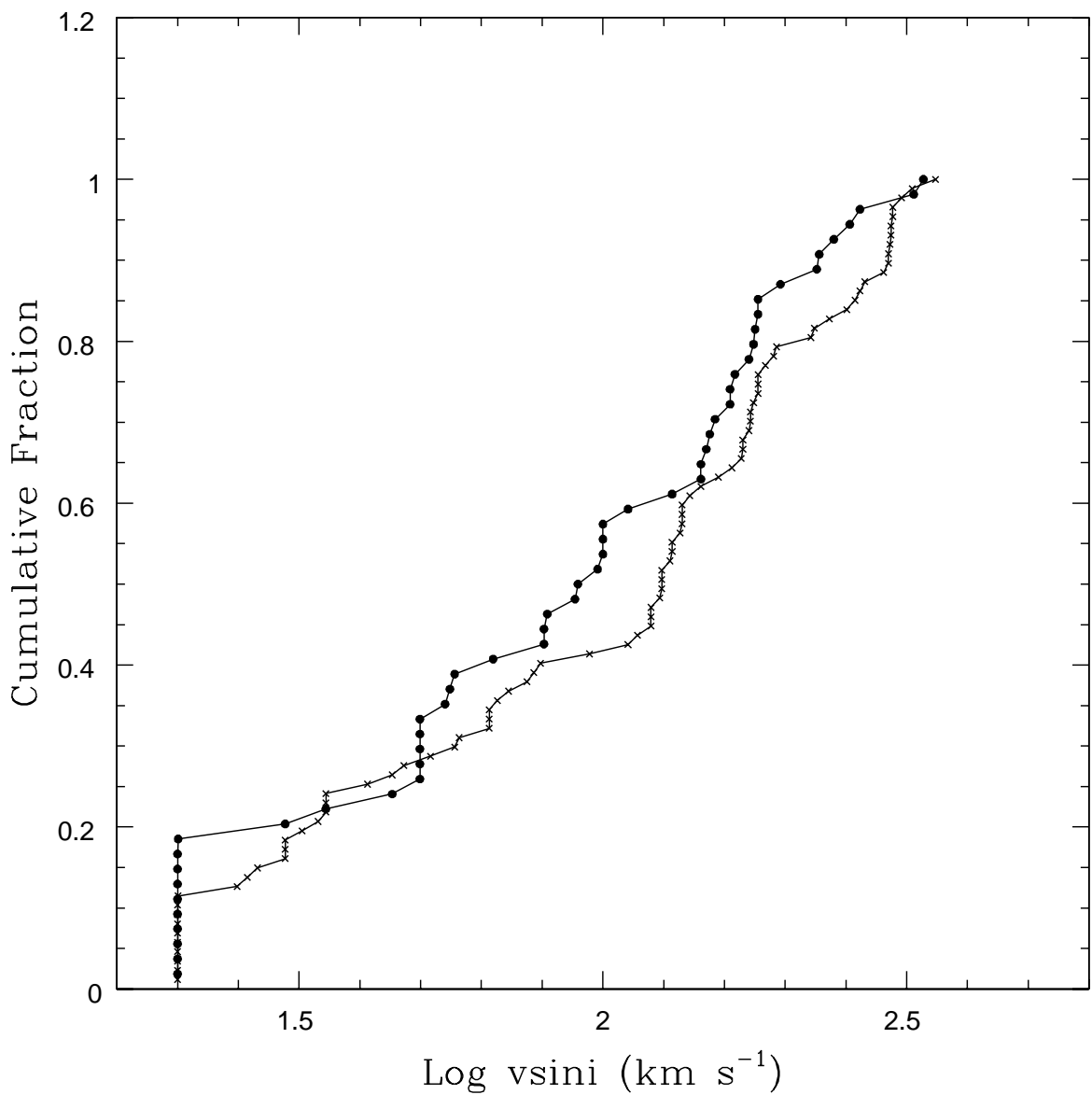}
\figcaption{Comparison of the cumulative distribution of {\vsini} for 54 stars in the young low density regions NGC 6823, Orion c, and Upper Sco (filled circles) with the distribution for 87 stars in the older low density regions I Lac, Upper CenLup, Lower Cen-Crux, and Ori a (crosses). A K-S test indicates that the probability that these two samples are drawn from the same parent population is 0.15.}
\end{figure}
\clearpage
\begin{figure}
\plotone{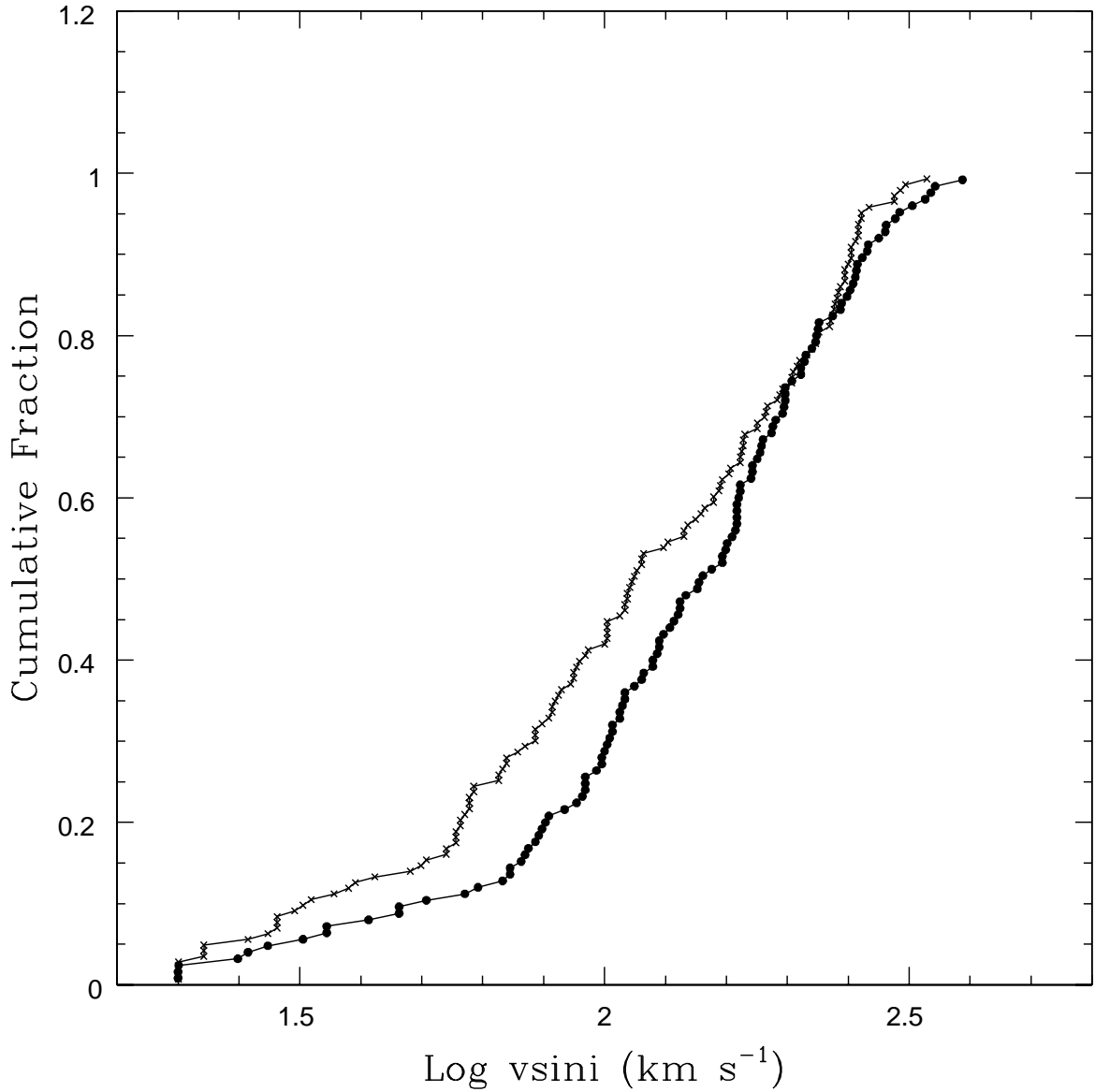}
\figcaption{Comparison of the cumulative distribution of {\vsini} for 125 stars in the young high density regions Ori b, NGC 7380, IC 1805, NGC 2244, NGC 6611, and Cyg OB2 (filled circles) with the distribution for 143 stars in the older clusters h and $\chi$ Per (crosses).  The probability that these two distributions are drawn from the same parent population is 0.02.}
\end{figure}
\clearpage
\begin{figure}
\plotone{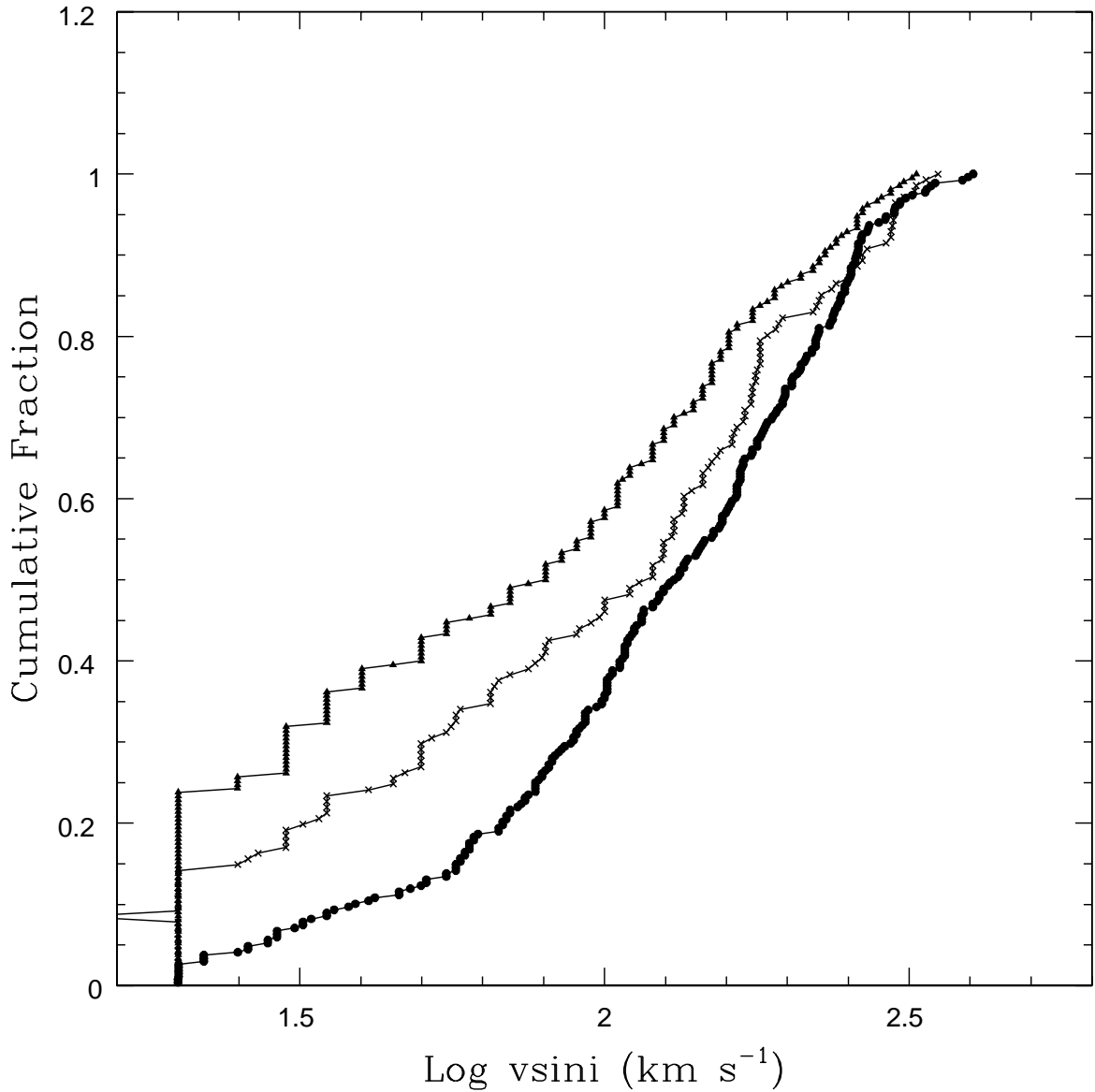}
\figcaption{Comparison of the cumulative distribution of {\vsini} for 141 stars in low density regions (crosses) with the distribution for 268 stars in high density regions (circles) and 810 field stars with luminosity classes IV and V (triangles).  A K-S test indicates that the probability that distribution for the stars in high density regions is drawn from the same sample as those in low density regions is 0.001; the probability that the distribution for the stars in the high density region is drawn from the same parent population as that for the field stars is less than 0.00005.}
\end{figure}
\clearpage
\begin{figure}
\plotone{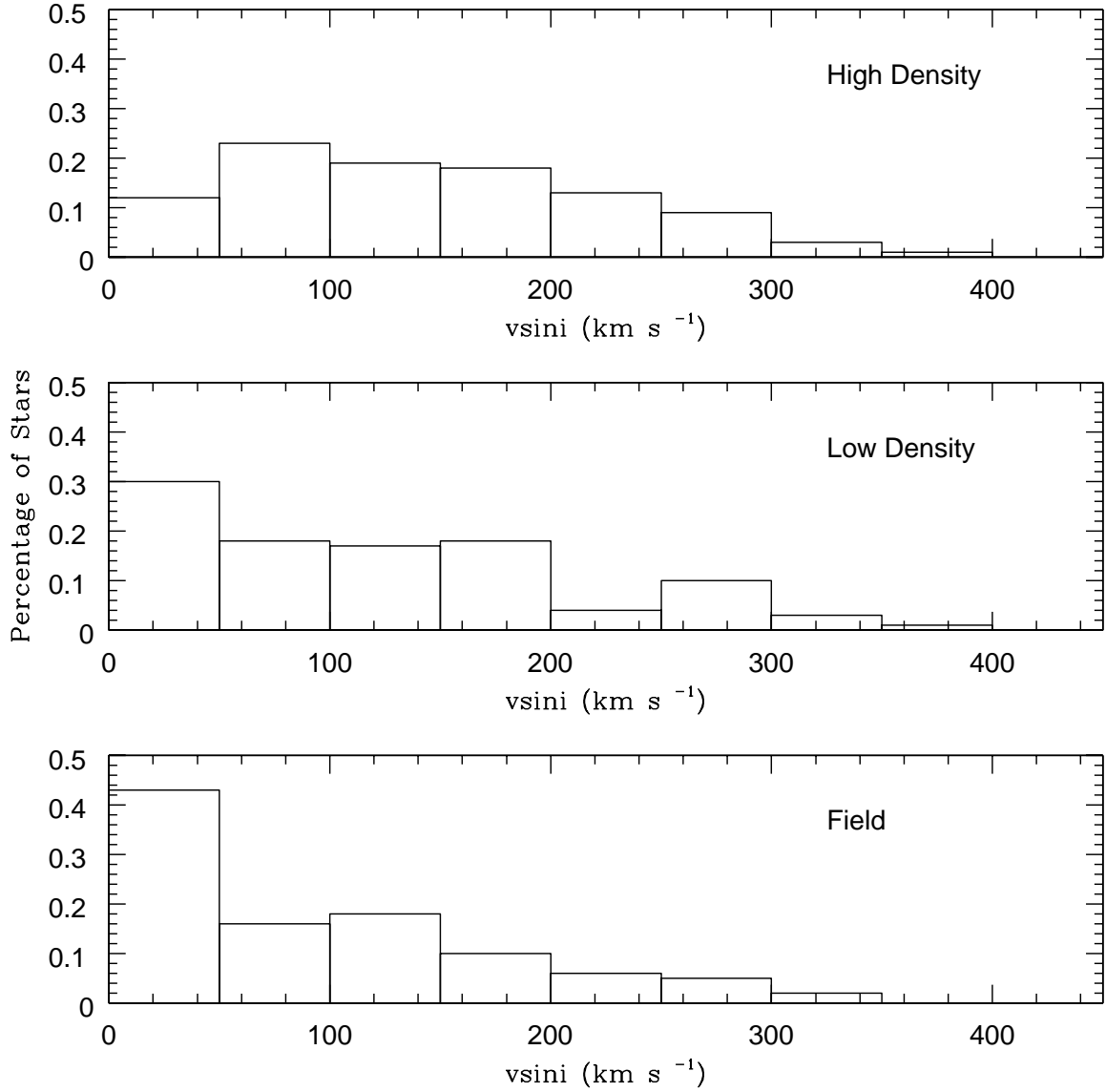}
\figcaption{Histograms showing the distribution of rotational velocities for stars in high density regions, low density regions and the field.}
\end{figure}
\clearpage

\clearpage
\begin{deluxetable}{ccccccccrc}
\tabletypesize{\scriptsize}




\tablecaption{Stellar Data}

\tablenum{1}

\tablehead{\colhead{Cluster} & \colhead{WEBDA} & \colhead{RA} & \colhead{Dec} & \colhead{Spectral Type} & \colhead{V} & \colhead{B-V} & \colhead{U-B} & \colhead{{\vsini}} & \colhead{Source} \\ 
\colhead{} & \colhead{} & \colhead{(J2000.0)} & \colhead{(J2000.0)} & \colhead{} & \colhead{} & \colhead{} & \colhead{} & \colhead{km s$^{-1}$} & \colhead{}} 

\startdata
IC 1805 & 72 & 02 30 56.92 & +61 16 29.8 & B2.5V & 12.39 & 0.63 & -0.25 & 259 & \nodata \\
IC 1805 & 82 & 02 31 19.75 & +61 30 16.2 & B2.5V & 12.51 & 0.83 & -0.17 & 191 & \nodata \\
IC 1805 & 94 & 02 31 48.47 & +61 34 55.8 & B2V & 13.93 & 0.76 & -0.08 & 203 & \nodata \\
IC 1805 & 109 & 02 32 06.48 & +61 29 54.2 & B1.5V & 13.93 & 0.91 & -0.06 & 164 & \nodata \\
IC 1805 & 121 & 02 32 18.38 & +61 27 52.8 & B1V & 11.61 & 0.66 & -0.31 &  90 & \nodata \\
IC 1805 & 130 & 02 32 29.93 & +61 27 07.6 & \nodata & 13.38 & 0.65 & -0.12 & 133 & 1 \\
IC 1805 & 136 & 02 32 34.56 & +61 32 19.1 & B1.5V & 11.05 & 0.60 & -0.34 &  79 & \nodata \\
IC 1805 & 143 & 02 32 40.85 & +61 27 59.8 & B1V & 11.38 & 0.51 & -0.45 & 320 & 1 \\
IC 1805 & 149 & 02 32 42.73 & +61 29 34.4 & B1.5V & 11.22 & 0.49 & -0.38 &  86 & 1 \\
IC 1805 & 156 & 02 32 46.05 & +61 27 56.8 & B1IV & 12.05 & 0.60 & -0.37 &  99 & \nodata  \\
IC 1805 & 157 & 02 32 46.99 & +61 31 32.1 & \nodata & 13.43 & 0.64 & -0.18 &  51 & 1 \\
IC 1805 & 158 & 02 32 47.57 & +61 27 00.0 & \nodata & 12.76 & 0.97 & -0.03 & 162 & 1 \\
IC 1805 & 161 & 02 32 55.24 & +61 38 56.9 & B1.5V & 10.90 & 0.42 & -0.39 & 174 & \nodata \\
IC 1805 & 166 & 02 32 57.83 & +61 27 26.6 & B2.5V & 11.99 & 0.57 & -0.28 & 107 & 1 \\
IC 1805 & 167 & 02 32 59.47 & +61 36 34.8 & B1.5V & 12.33 & 0.45 & -0.33 & 282 & \nodata \\
IC 1805 & 169 & 02 32 58.99 & +61 22 23.5 & B1.5V & 11.73 & 0.66 & -0.27 &  93 & \nodata \\
IC 1805 & 174 & 02 33 04.75 & +61 28 21.1 & B2.5V & 11.58 & 0.55 & -0.34 & 197 & \nodata \\
IC 1805 & 180 & 02 33 09.06 & +61 27 46.1 & B1V & 12.93 & 0.65 & -0.19 & 336 & \nodata \\
IC 1805 & 188 & 02 33 18.24 & +61 32 12.0 & --- & 12.65 & 0.53 & -0.26 &  28 & 1 \\
IC 1805 & 191 & 02 33 20.50 & +61 32 23.3 & B2.5V & 12.97 & 0.62 & -0.25 &  78 & \nodata \\
IC 1805 & 211 & 02 33 43.01 & +61 26 12.2 & B1V & 11.08 & 0.57 & -0.37 & 198 & \nodata \\
IC 1805 & 225 & 02 33 53.30 & +61 18 26.6 & B2V & 13.70 & 0.79 & -0.13 & 253 & \nodata \\
IC 1805 & 243 & 02 34 15.02 & +61 24 40.4 & B1V & 11.20 & 0.54 & -0.41 & 250 & \nodata \\
IC 1805 & 260 & 02 34 31.48 & +61 30 35.2 & B2V & 11.55 & 0.45 & -0.35 &  93 & \nodata \\
IC 1805 & 277 & 02 34 47.66 & +61 26 17.1 & \nodata & 12.88 & 0.66 & -0.19 & 156 & 1 \\
IC 1805 & 288 & 02 35 05.28 & +61 28 09.9 & B1V & 11.10 & 0.50 & -0.41 & 198 & \nodata \\
IC 1805 & 1329 & 02 34 21.77 & +61 48 37.5 & \nodata & 12.69 & 0.54 & -0.34 &  68 & 1 \\
IC 1805 & 1824 & 02 36 03.98 & +61 25 07.2 & \nodata & 12.98 & 0.58 & -0.24 & 289 & 1 \\
IC 1805 & 1924 & 02 30 48.58 & +61 17 16.8 & B2.5V & 12.56 & 0.62 & -0.24 & 224 & \nodata \\
IC 1805 & 1930 & 02 30 39.11 & +61 21 05.5 & \nodata & 12.60 & 0.84 & 0.00 &  35 & 1 \\
IC 1805 & 1939 & 02 30 29.49 & +61 19 44.8 & \nodata & 12.83 & 0.79 & -0.09 &  41 & 1 \\
IC 1805 & 1940 & 02 30 28.84 & +61 32 56.2 & B2.5V & 13.28 & 1.07 & 0.06 &  97 & \nodata \\
IC 1805 & 1941 & 02 30 26.34 & +61 27 43.3 & B2.5V & 12.81 & 1.09 & 0.06 & 156 & \nodata \\
IC 1805 & 1947 & 02 30 15.41 & +61 23 42.2 & B2.5V & 13.70 & 0.94 & -0.06 & 387 & \nodata \\
IC 1805 & 1984 & 02 29 27.89 & +61 32 22.0 & B2.5V & 12.71 & 0.78 & -0.12 & 106 & \nodata \\
 \\
NGC 2244 & 62 & 06 31 25.44 & +05 02 09.6 & \nodata & 12.93 & 0.71 & -0.14 &  11 & 1 \\
NGC 2244 & 79 & 06 31 31.48 & +04 51 00.0 & B2.5V & 10.70 & 0.16 & -0.55 & 128 & \nodata  \\
NGC 2244 & 80 & 06 31 33.48 & +04 50 40.0 & B1V & 9.39 & 0.14 & -0.64 & 237 & \nodata \\
NGC 2244 & 115 & 06 31 38.40 & +05 01 36.6 & B0.2V & 8.03 & 0.08 & -0.75 & 159 & \nodata \\
NGC 2244 & 119 & 06 31 55.12 & +04 57 19.6 & \nodata & 12.35 & 0.33 & -0.26 &   4 & 1 \\
NGC 2244 & 123 & 06 31 55.25 & +04 56 22.0 & \nodata & 11.79 & 0.23 & -0.36 & 175 & 1 \\
NGC 2244 & 128 & 06 31 52.02 & +04 55 57.5 & B1V & 9.39 & 0.14 & -0.68 & 178 & \nodata \\
NGC 2244 & 130 & 06 31 47.90 & +04 54 18.3 & \nodata & 11.65 & 0.26 & -0.43 &  77 & 1 \\
NGC 2244 & 167 & 06 32 02.59 & +05 05 08.9 & B2V & 10.70 & 0.19 & -0.59 & 116 & \nodata \\
NGC 2244 & 172 & 06 32 09.84 & +05 02 13.6 & \nodata & 11.27 & 0.27 & -0.44 & 260 & 1 \\
NGC 2244 & 192 & 06 32 09.64 & +04 55 57.1 & \nodata & 12.55 & 0.45 & -0.14 &  70 & 1 \\
NGC 2244 & 193 & 06 31 58.95 & +04 55 40.1 & B1.5V & 10.36 & 0.21 & -0.54 & 100 & \nodata \\
NGC 2244 & 194 & 06 32 15.49 & +04 55 20.5 & \nodata & 12.02 & 0.31 & -0.28 & 343 & 1 \\
NGC 2244 & 200A & 06 32 00.62 & +04 52 41.1 & \nodata & 8.54 & 0.14 & -0.70 &  74 & \nodata \\
NGC 2244 & 200B & 06 32 00.62 & +04 52 41.1 & \nodata & \nodata & \nodata & \nodata & 132 & \nodata \\
NGC 2244 & 201 & 06 32 06.15 & +04 52 15.6 & B1III & 9.74 & 0.15 & -0.68 &  46 & \nodata \\
NGC 2244 & 206 & 06 32 13.44 & +04 47 37.1 & \nodata & 12.04 & 0.54 & -0.14 & 305 & 1 \\
NGC 2244 & 253 & 06 32 29.40 & +04 56 56.3 & \nodata & 10.78 & 0.30 & -0.29 & 167 & 1 \\
NGC 2244 & 274 & 06 32 24.24 & +04 47 04.0 & B2.5V & 11.43 & 0.28 & -0.49 &  46 & \nodata \\
NGC 2244 & 279 & 06 32 34.96 & +04 44 39.5 & Be & 11.29 & 0.41 & -0.47 & 395 & \nodata \\
NGC 2244 & 353 & 06 32 59.38 & +04 56 22.7 & \nodata & 9.63 & 0.22 & -0.50 & 222 & 1 \\
NGC 2244 & 376 & 06 30 33.33 & +04 41 27.9 & \nodata & 9.71 & 0.54 & -0.47 &  26 & 1 \\
NGC 2244 & 391 & 06 33 43.99 & +04 45 55.5 & \nodata & 10.23 & 0.27 & -0.32 & 349 & 1 \\
NGC 2244 & 392 & 06 33 50.56 & +05 01 37.8 & B2.5V & 11.19 & 0.42 & -0.39 & 115 & \nodata \\
NGC 2244 & 1006 & 06 33 37.51 & +04 48 47.0 & B0.5V & 11.89 & 0.93 & -0.10 & 143 & \nodata \\
NGC 2244 & 1147 & 06 34 13.59 & +04 44 12.0 & \nodata & 9.88 & 0.26 & -0.45 & 256 & 1 \\
NGC 2244 & 1262 & 06 33 09.90 & +04 32 54.7 & \nodata & 12.34 & 0.45 & -0.16 & 108 & 1 \\
NGC 2244 & 1479 & 06 33 10.16 & +04 59 50.2 & B3V & 14.98 & 0.78 & -0.14 &  92 & \nodata \\
NGC 2244 & 1553 & 06 31 37.10 & +04 45 53.7 & B0.5V & 15.15 & 0.98 & 0.02 & 189 & \nodata \\
NGC 2244 & 1607 & 06 33 09.39 & +05 07 53.7 & \nodata & 10.61 & 0.22 & -0.45 & 136 & 1 \\
 \\
NGC 6611 & 207 & 18 18 36.80 & -13 47 33.3 & B1Ve & 12.07 & 0.53 & -0.28 & 103 & \nodata \\
NGC 6611 & 227 & 18 18 38.45 & -13 47 09.2 & B1.5Ve & 12.85 & 0.59 & -0.26 & 123 & \nodata \\
NGC 6611 & 231 & 18 18 38.52 & -13 45 56.4 & B1V & 12.71 & 0.75 & -0.26 & 196 & \nodata \\
NGC 6611 & 239 & 18 18 40.06 & -13 54 33.7 & B1V & 11.48 & 0.36 & -0.42 & 102 & \nodata \\
NGC 6611 & 254 & 18 18 40.82 & -13 46 52.3 & B1V & 10.80 & 0.47 & -0.43 &  32 & 2 \\
NGC 6611 & 267 & 18 18 41.75 & -13 46 44.0 & \nodata & 13.13 & 0.52 & -0.22 & 125 & 2 \\
NGC 6611 & 269 & 18 18 41.64 & -13 42 48.0 & B1.5V & 13.98 & 0.93 & -0.04 & 290 & \nodata \\
NGC 6611 & 289 & 18 18 44.15 & -13 48 56.7 & \nodata & 12.60 & 0.50 & -0.21 & 219 & \nodata \\
NGC 6611 & 290 & 18 18 44.91 & -13 56 22.5 & \nodata & 12.14 & 0.39 & -0.29 & 223 & \nodata \\
NGC 6611 & 297 & 18 18 44.59 & -13 45 48.3 & B1.5:V & 12.88 & 0.67 & -0.21 & 270 & 2 \\
NGC 6611 & 300 & 18 18 45.10 & -13 47 47.3 & B1.5Ve & 12.69 & 0.52 & -0.24 & 271 & \nodata \\
NGC 6611 & 301 & 18 18 45.04 & -13 46 25.0 & B2V & 12.22 & 0.57 & -0.29 & 120 & 2 \\
NGC 6611 & 306 & 18 18 45.09 & -13 45 41.1 & B1.5Ve & 12.77 & 0.68 & -0.22 & 258 & \nodata \\
NGC 6611 & 311 & 18 18 45.64 & -13 47 53.4 & B2.5Ve & 13.10 & 0.51 & -0.21 & 165 & 2 \\
NGC 6611 & 336 & 18 18 49.23 & -13 48 04.4 & \nodata & 13.29 & 0.52 & -0.20 & 265 & 2 \\
NGC 6611 & 343 & 18 18 49.44 & -13 46 50.2 & B1V & 11.72 & 0.85 & -0.17 & 300 & \nodata \\
NGC 6611 & 409 & 18 18 57.43 & -13 52 12.5 & \nodata & 12.84 & 0.40 & -0.24 & 214 & \nodata \\
NGC 6611 & 444 & 18 19 00.49 & -13 42 41.1 & B1.5V & 12.74 & 0.81 & -0.17 & 133 & \nodata \\
NGC 6611 & 483 & 18 19 06.57 & -13 43 30.5 & \nodata & 10.99 & 0.41 & -0.20 & 165 & 2 \\
NGC 6611 & 536 & 18 19 18.54 & -13 55 40.3 & B1.5V & 11.46 & 0.22 & -0.36 &  73 & \nodata \\
NGC 6611 & 541 & 18 19 19.19 & -13 43 52.3 & B2.5V & 13.31 & 0.60 & -0.22 &  70 & 2 \\
NGC 6611 & 567 & 18 18 16.87 & -13 58 46.6 & \nodata & 11.99 & 0.36 & -0.31 & 245 & \nodata \\
NGC 6611 & 587 & 18 18 56.73 & -13 59 48.9 & \nodata & 11.97 & 0.41 & -0.27 & 165 & \nodata \\
NGC 6611 & 597 & 18 19 13.17 & -13 57 38.5 & \nodata & 12.30 & 0.39 & -0.30 & 210 & \nodata \\
NGC 6611 & 601 & 18 19 20.09 & -13 54 21.9 & B1.5V & 10.68 & 0.36 & -0.51 & 198 & \nodata \\
NGC 6611 & 607 & 18 19 32.88 & -13 55 50.9 & \nodata & 12.48 & 0.47 & -0.23 &  80 & \nodata \\
NGC 6611 & \nodata & 18 20 01.45 & -13 53 30.6 & \nodata & 12.23 & 0.50 & -0.15 &  99 & \nodata \\
 \\
NGC 6823 & 67 & 19 43 16.90 & +23 19 11.9 & B1V & 12.68 & 0.80 & -0.25 &  57 & \nodata \\
NGC 6823 & 91 & 19 43 36.62 & +23 21 08.3 & B1V & 10.95 & 0.45 & -0.47 & 162 & \nodata \\
NGC 6823 & 364 & 19 43 32.52 & +23 22 10.7 & B1.5III & 10.69 & 0.35 & -0.55 &  66 & \nodata \\
NGC 6823 & \nodata & 19 45 06.12 & +23 58 37.3 & B1V & 11.19 & 0.68 & -0.28 & 337 & \nodata \\
NGC 6823 & A & 19 45 47.54 & +24 06 00.4 & B1V & 10.91 & 0.69 & -0.26 & 110 & \nodata \\
NGC 6823 & B & 19 45 47.54 & +24 06 00.4 & \nodata & \nodata & \nodata & \nodata & 153 & \nodata \\
NGC 6823 & \nodata & 19 44 49.17 & +24 01 34.6 & B2.5V & 12.87 & 0.82 & -0.10 &  81 & \nodata \\
NGC 6823 & \nodata & 19 45 18.34 & +24 00 59.7 & B2V & 12.10 & 0.69 & -0.23 & 177 & \nodata \\
 \\
Cyg OB2 & 169 & 20 31 56.27 & +41 33 05.3 & B1.5V & 13.90 & 1.21 & 0.27 & 103 & \nodata \\
Cyg OB2 & 174 & 20 31 56.90 & +41 31 48.0 & B1.5V & 12.55 & 1.21 & 0.23 &  59 & \nodata \\
Cyg OB2 & 187 & 20 32 03.74 & +41 25 10.9 & B0.5V & 13.24 & 1.52 & 0.39 & 101 & \nodata \\
Cyg OB2 & 250 & 20 32 26.10 & +41 29 39.0 & B1V & 12.88 & 1.06 & 0.16 &  62 & \nodata \\
Cyg OB2 & 292 & 20 32 37.03 & +41 23 05.1 & B1V & 13.08 & 1.51 & 0.42 & 108 & \nodata \\
Cyg OB2 & 378 & 20 32 59.61 & +41 15 14.6 & B0V & 13.49 & 2.10 & 0.95 & 122 & \nodata \\
Cyg OB2 & 403 & 20 33 05.55 & +41 43 37.2 & B1.5V & 12.94 & 1.49 & 0.47 & 106 & \nodata  \\
Cyg OB2 & 429 & 20 33 10.50 & +41 22 22.8 & B0V & 12.98 & 1.56 & 0.43 & 130 & \nodata \\
Cyg OB2 & 515 & 20 33 23.24 & +41 13 41.9 & B1V & 14.66 & 2.03 & 0.99 & 188 & \nodata \\
Cyg OB2 & 588 & 20 33 37.02 & +41 16 11.4 & B0V & 12.40 & 1.66 & 0.40 & 244 & \nodata \\
Cyg OB2 & 605 & 20 33 39.84 & +41 22 52.4 & B0.5V & 11.78 & 1.19 & 0.20 & 142 & \nodata \\
Cyg OB2 & 692 & 20 33 59.32 & +41 05 38.4 & B0V & 13.61 & 1.69 & 0.46 & 158 & \nodata \\
Cyg OB2 & 793 & 20 34 43.51 & +41 29 04.8 & B1.5III & 12.29 & 1.54 & 0.30 & 182 & \nodata \\
 \\
NGC 7380 & 8 & 22 47 12.57 & +58 08 41.1 & B0.5V & 10.62 & 0.26 & -0.58 &  93 & \nodata \\
NGC 7380 & 9 & 22 47 39.23 & +58 09 32.4 & B0.5V & 10.66 & 0.35 & -0.50 & 150 & \nodata \\
NGC 7380 & 34 & 22 47 35.04 & +58 07 36.3 & B1.5V & 11.81 & 0.37 & -0.45 & 123 & \nodata \\
NGC 7380 & 35 & 22 47 08.31 & +58 04 45.2 & B1.5V & 11.89 & 0.32 & -0.47 & 213 & \nodata \\
NGC 7380 & 42 & 22 47 47.15 & +58 03 07.3 & B1V & 12.28 & 0.60 & -0.29 & 225 & \nodata \\
NGC 7380 & 136 & 22 48 16.27 & +58 00 47.5 & B0.5V & 10.41 & 0.51 & -0.42 & 181 & \nodata \\
NGC 7380 & 138 & 22 47 04.89 & +58 06 01.9 & B1V & 11.25 & 0.35 & -0.49 & 167 & \nodata \\
NGC 7380 & 142 & 22 47 45.66 & +58 06 48.6 & B1V & 11.77 & 0.50 & -0.41 & 166 & \nodata \\
NGC 7380 & 2326 & 22 46 31.42 & +58 01 59.4 & B1.5V & 14.23 & 0.84 & -0.11 & 112 & \nodata \\
NGC 7380 & 5755 & 22 49 43.46 & +58 11 04.8 & B1III & 10.61 & 0.26 & -0.59 &  81 & \nodata \\
\enddata


\tablecomments{Identification numbers for Cyg OB2 are from Massey and Thompson (1991)}
\tablerefs{(1)Huang \& Gies (2006a); (2) Dufton et al. (2006)}
\end{deluxetable}
\clearpage

\begin{deluxetable}{ccrccrccr}
\tabletypesize{\scriptsize}
\tablecaption{Data for Stellar Associations}
\tablewidth{0pt}
\tablenum{2}

\tablehead{\colhead{Association} & \colhead{HD} & \colhead{vsini} & \colhead{Association} & \colhead{HD} & \colhead{vsini} & \colhead{Association} & \colhead{HD} & \colhead{{\vsini}} \\ 
\colhead{} & \colhead{} & \colhead{(km s$^{-1}$)} & \colhead{} & \colhead{} & \colhead{(km s$^{-1}$)} & \colhead{} & \colhead{} & \colhead{(km s$^{-1}$)} } 

\startdata
I Lac & 209961 & 145 & UCL & 120307 & 65 & Orion b & 36485 & 35 \\
I Lac & 212883 & 5 & UCL & 120324 & 130 & Orion b & 36646 & 20 \\
I Lac & 212978 & 120 & UCL & 121743 & 79 & Orion b & 36695 & 120 \\
I Lac & 213420 & 70 & UCL & 121790 & 124 & Orion b & 36779 & 175 \\
I Lac & 213976 & 135 & UCL & 122980 & 15 & Orion b & 36954 & 180 \\
I Lac & 214167 & 265 & UCL & 124367 & 323 & Orion b & 37479 & 165 \\
I Lac & 214263 & 125 & UCL & 125823 & 15 & Orion b & 37744 & 25 \\
I Lac & 214432 & 185 & UCL & 129056 & 16 & Orion b & 37756 & 75 \\
I Lac & 214993 & 30 & UCL & 129116 & 129 & Orion b & 37776 & 145 \\
I Lac & 215191 & 180 & UCL & 130807 & 27 & Orion b & 37903 & 210 \\
I Lac & 215227 & 30 & UCL & 131120 & 57 & Orion c & 33328 & 325 \\
I Lac & 216684 & 125 & UCL & 132200 & 32 & Orion c & 35337 & 15 \\
I Lac & 216851 & 310 & UCL & 133955 & 135 & Orion c & 36151 & 50 \\
I Lac & 216916 & 120 & UCL & 134687 & 13 & Orion c & 36285 & 15 \\
I Lac & 217101 & 130 & UCL & 136298 & 193 & Orion c & 36430 & 15 \\
I Lac & 217227 & 30 & UCL & 136504 & 41 & Orion c & 36629 & 5 \\
I Lac & 218344 & 95 & UCL & 136664 & 177 & Orion c & 36936 & 180 \\
USco & 141637 & 227 & UCL & 137432 & 77 & Orion c & 36958 & 50 \\
USco & 142114 & 240 & UCL & 138690 & 270 & Orion c & 36959 & 5 \\
USco & 142184 & 255 & UCL & 138769 & 67 & Orion c & 36981 & 145 \\
USco & 142378 & 225 & UCL & 139365 & 134 & Orion c & 37000 & 80 \\
USco & 142669 & 98 & UCL & 140008 & 11 & Orion c & 37016 & 100 \\
USco & 142883 & 14 & UCL & 143118 & 191 & Orion c & 37017 & 165 \\
USco & 142990 & 178 & UCL & 143699 & 170 & Orion c & 37040 & 145 \\
USco & 143018 & 100 & UCL & 144294 & 252 & Orion c & 37055 & 50 \\
USco & 144217 & 91 & UCL & 151985 & 52 & Orion c & 37058 & 5 \\
USco & 144334 & 55 & Orion a & 31331 & 180 & Orion c & 37129 & 50 \\
USco & 144470 & 100 & Orion a & 34511 & 35 & Orion c & 37209 & 35 \\
USco & 144661 & 45 & Orion a & 34748 & 295 & Orion c & 37303 & 265 \\
USco & 145482 & 174 & Orion a & 35007 & 35 & Orion c & 37356 & 10 \\
USco & 145502 & 162 & Orion a & 35039 & 5 & Orion c & 37481 & 90 \\
USco & 145792 & 30 & Orion a & 35079 & 174 & Orion c & 37526 & 130 \\
USco & 147165 & 56 & Orion a & 35148 & 300 & Orion c & 37807 & 10 \\
USco & 147701 & 80 & Orion a & 35149 & 220 & Orion c & 39291 & 150 \\
USco & 147888 & 180 & Orion a & 35298 & 260 & Orion c & 39777 & 20 \\
USco & 147933 & 196 & Orion a & 35299 & 0 & Orion c & 294264 & 50 \\
USco & 148184 & 148 & Orion a & 35407 & 295 & \nodata & \nodata & \nodata \\
LCC & 98718 & 353 & Orion a & 35411 & 35 & \nodata & \nodata & \nodata \\
LCC & 103079 & 47 & Orion a & 35502 & 290 & \nodata & \nodata & \nodata \\
LCC & 105382 & 75 & Orion a & 35575 & 120 & \nodata & \nodata & \nodata \\
LCC & 106490 & 135 & Orion a & 35588 & 170 & \nodata & \nodata & \nodata \\
LCC & 106983 & 65 & Orion a & 35715 & 110 & \nodata & \nodata & \nodata \\
LCC & 108257 & 298 & Orion a & 35730 & 58 & \nodata & \nodata & \nodata \\
LCC & 108483 & 169 & Orion a & 35762 & 163 & \nodata & \nodata & \nodata \\
LCC & 109668 & 114 & Orion a & 35777 & 300 & \nodata & \nodata & \nodata \\
LCC & 110879 & 139 & Orion a & 35792 & 65 & \nodata & \nodata & \nodata \\
LCC & 110956 & 26 & Orion a & 35912 & 5 & \nodata & \nodata & \nodata \\
LCC & 112078 & 298 & Orion a & 36013 & 297 & \nodata & \nodata & \nodata \\
LCC & 112092 & 34 & Orion a & 36133 & 236 & \nodata & \nodata & \nodata \\
LCC & 113791 & 25 & Orion a & 36166 & 125 & \nodata & \nodata & \nodata \\
LCC & 116087 & 223 & Orion a & 36267 & 155 & \nodata & \nodata & \nodata \\
\nodata & \nodata & \nodata & Orion a & 36351 & 20 & \nodata & \nodata & \nodata \\
\nodata & \nodata & \nodata & Orion a & 36392 & 45 & \nodata & \nodata & \nodata \\
\nodata & \nodata & \nodata & Orion a & 36741 & 175 & \nodata & \nodata & \nodata \\
\nodata & \nodata & \nodata & Orion a & 36824 & 175 & \nodata & \nodata & \nodata \\
\nodata & \nodata & \nodata & Orion a & 37490 & 180 & \nodata & \nodata & \nodata \\
\enddata



\end{deluxetable}

\clearpage
\begin{deluxetable}{lccccccc}
\tabletypesize{\scriptsize}
\tablecaption{Properties of Clusters and Associations}

\tablenum{3}

\tablehead{\colhead{Identification} & \colhead{Distance} & \colhead{Age} & \colhead{Early B Stars} & \colhead{Median Radius} & \colhead{Median Radius} & \colhead{Mass} & \colhead{Density} \\ 
\colhead{} & \colhead{(pc)} & \colhead{(Myr)} & \colhead{Number} & \colhead{(Degrees)} & \colhead{(pc)} & \colhead{(M$_{\sun}$)} & \colhead{(M$_{\sun}$ pc $^{-3}$)} } 

\startdata
NGC 6823 & 2300 & 2-7 & 60 & 0.85 & 34.1 & 8700 & 0.05 \\
I Lac & 368 & 12-16 & 19 & 3.23 & 20.7 & 2760 & 0.07 \\
Upper Cen-Lup & 142 & 14-15 & 29 & 8.9 & 22.1 & 4200 & 0.09 \\
Lower Cen-Crux & 118 & 11-12 & 14 & 7.3 & 15.0 & 2030 & 0.14 \\
Upper Sco & 145 & 5-6 & 21 & 5.6 & 14.2 & 3050 & 0.26 \\
Orion a & 380 & 11.4 & 38 & 2.5 & 16.6 & 5520 & 0.29 \\
Orion c & 398 & 4.6 & 40 & 1.8 & 12.5 & 5810 & 0.71 \\
Orion b & 363 & 1.7 & 26 & 1 & 6.3 & 3770 & 3.55 \\
NGC 7380 & 3730 & 2 & 42 & 0.1 & 6.5 & 6100 & 5.3 \\
IC 1805 & 2345 & 1-3 & 99 & 0.17 & 7.1 & 14400 & 9.6 \\
NGC 2244 & 1880 & 1-3 & 54 & 0.17 & 5.6 & 7840 & 10.8 \\
NGC 6611 & 1995 & 1-5 & 170 & 0.17 & 5.9 & 24700 & 28.5 \\
Cyg OB2 & 1740 & 1-4 & 160 & 0.17 & 5.2 & 23200 & 40.4 \\
${\chi}$ Per & 2345 & 12.8 & 78 & 0.06 & 2.5 & 11300 & 178 \\
h Per & 2345 & 12.8 & 110 & 0.05 & 2.1 & 16000 & 416 \\
\enddata




\end{deluxetable}

\end{document}